\documentclass[preprint, superscriptaddress, aps, prl]{revtex4-2}
\usepackage{graphicx}
\usepackage{bm,amsmath,amssymb}
\usepackage[version=4]{mhchem}
\usepackage[colorlinks=true,
            linkcolor=blue,
            citecolor=blue,
            urlcolor=blue
]{hyperref}

\begin{document}

% ----------------------------------------
% 		Title and Authors section
% ----------------------------------------
\title{
	Unraveling Shear Strain Induced Ferroelectric-to-Antiferroelectric Phase Transition and Accessing Intrinsic Antiferroelectricity in Two-dimensional NbOCl\texorpdfstring{\textsubscript{2}}{2}
}

\author{Jiawei Mao}
\affiliation{Frontier Institute of Science and Technology, State Key Laboratory of Electrical Insulation and Power Equipment, Xi’an Jiaotong University, Xi’an 710049, China}

\author{Yinglu Jia}
\affiliation{Department of Chemical and Nano Engineering, University of California San Diego, La Jolla, CA 92093, USA}

\author{Gaoyang Gou}
\thanks{Contact author: gougaoyang@mail.xjtu.edu.cn}
\affiliation{Frontier Institute of Science and Technology, State Key Laboratory of Electrical Insulation and Power Equipment, Xi’an Jiaotong University, Xi’an 710049, China}

\author{Shi Liu}
\thanks{Contact author: liushi@westlake.edu.cn}
\affiliation{Department of Physics, School of Science and Research Center for Industries of the Future, Westlake University, Hangzhou, 310030, China}
\affiliation{Institute of Natural Sciences, Westlake Institute for Advanced Study, Hangzhou, Zhejiang 310024, China}

\author{Xiao Cheng Zeng}
\thanks{Contact author: xzeng26@cityu.edu.hk}
\affiliation{Department of Materials Science and Engineering, City University of Hong Kong, Hong Kong 999077, China}
\date{\today}

% ----------------------------------------
% 			Abstract section
% ----------------------------------------
\begin{abstract}
	Compared to the well studied two-dimensional (2D) ferroelectricity, much rare is the appearance of 2D antiferroelectricity, where local dipoles from the nonequivalent sublattices within 2D monolayers are oppositely orientated. Using \ce{NbOCl2} monolayer with competing ferroelectric (FE) and antiferroelectric (AFE) phases as a 2D material platform, we demonstrate the emerging of intrinsic antiferroelectricity in \ce{NbOCl2} monolayer under the experimentally accessible shear strain, and new functionality associated with electric field induced AFE-to-FE phase transition. Specifically, the complex configuration space accommodating FE and AFE phases, polarization switching kinetics and finite temperature thermodynamic properties of 2D \ce{NbOCl2}, are all accurately predicted by large-scale molecular dynamic (MD) simulations based on deep learning interatomic potential (DP) model. Moreover, room temperature stable antiferroelectricity with low polarization switching barrier and one-dimensional (1D) collinear polarization arrangement is predicted in shear deformed \ce{NbOCl2} monolayer. Transition from AFE to FE phase in 2D \ce{NbOCl2} can be triggered by the low critical electric field, leading to the double polarization-electric ($P$-$E$) loop with small hysteresis. A new type optoelectronic device composed of AFE-\ce{NbOCl2}, enabling electric "writing" and nonlinear optical "reading" logical operation with fast operation speed and low power consumption is also proposed.
\end{abstract}
\maketitle
\clearpage

Spontaneous symmetry breaking in ferroic materials with the specific order parameters can activate the unique material properties and functional responses \cite{Nat_Mater_6_21_2007}. Compared to FE materials with spontaneous ferroelectricity and breaking of spatial inversion symmetry, AFE materials posses the rich structural phase transition features and new application prospects, such as thermal switching \cite{Nat_Mater_23_944_2024}, dielectric energy storage \cite{Adv_Mater._34_2105967_2022}, electrocaloric \cite{Science_311_1270_2006} and electrostrictive effects \cite{Adv_Mater_21_4716_2009} that are less accessible in the FE counterparts \cite{Prog_Mater_Sci_142_101231_2024}. However, antiferroelectricity has a more stringent structural and energetic requirement than ferroelectricity. Specifically, the compensating net polarizations in AFE materials originate from the opposite dipole arrangement from the nonequivalent sublattices, which are usually accompanied by the unit-cell doubling or structural modulation \cite{Prog_Mater_Sci_142_101231_2024}. Moreover, the presence of low-energy FE phase in proximity to the stable AFE phase is also required. The competing AFE and FE phases are both associated with the structural instabilities inherent from a same high-symmetry reference structure \cite{KarinRabe}. Most importantly, electric field induced transition from AFE to FE phase is the key aspect of the unique functional responses and application potential for antiferroelectricity \cite{Phys_Rev_B_90_140103R_2014}.

The stringent requirement of antiferroelectricity also applies for 2D van der Waals (vdW) layered materials, where intrinsic 2D antiferroelectricity is much rarer than 2D ferroelectricity. In fact, ferroelectricity are widely reported and extensively investigated in various 2D FE materials with displacive ferroelectricity (e.g. \ce{CuInP2S6} \cite{Nat_Commun_7_12357_2016}, SnTe \cite{Science_353_274_2016} and $\alpha$-\ce{In2Se3} \cite{Nat_Commun_8_14956_2017,PRL_120_227601_2018}, etc) and sliding ferroelectricity (e.g. 1T'-\ce{WTe2} \cite{Nature_560_336_2018}, 1T'-\ce{ReS2} \cite{PRL_128_067601_2022}, 3R-\ce{MoS2} \cite{Science_385_57_2024}, 3R-BN \cite{Nature_629_74_2024}, $\gamma$-GaSe \cite{Nat_Commun_14_2757_2023}, and $\gamma$-InSe \cite{Nat_Commun_14_36_2023}, etc). While antiferroelectricity are reported in very limited cases that are often termed as "multi-polar" 2D materials \cite{Nano_Lett_22_4083_2022}. For example, in the experimentally synthesized 2D AFE materials such as $\alpha$-GeSe \cite{ACS_Nano_16_1308_2022} and \ce{CuBiP2Se6} \cite{Adv_Mater_37_2419204_2025}, each of their monolayer has the net FE polarization, antiferroelectricity appears in those even-numbered multilayers with antiparallel monolayer-polarization arrangement. In this context, intrinsic 2D antiferroelectricity able to persist in monolayer limit (breaking even-layer-number requirement) has rarely been realized.

Combining the high-throughput screening of vdW layered materials database and first-principles calculations, we identified \ce{NbO\textit{X}2} (\textit{X} = Cl, Br and I) as a new group of 2D FE materials \cite{Jia2019}. Especially, owing to its stable in-plane ferroelectricity, large polarization magnitude \cite{ACSNano2023} and giant second harmonic generation (SHG) response \cite{Nature_613_53_2023,PRL_132_246902_2024}, \ce{NbOCl2} becomes an ideal system to explore 2D ferroelectricity and second-order optical nonlinearity. Beside, unit-cell doubling induced by Peierls distortion between $\text{Nb}^{4+}$ cations of \ce{NbOCl2} naturally meets the structural requirement of antiferroelectricity. Moreover, the metastable AFE phase in proximity to stable FE phase is indeed predicted for 2D \ce{NbOCl2} monolayer and multilayer \cite{Jia2019}. Alternating the energy sequence between the competing FE and AFE phases by a certain control parameter \cite{KarinRabe}, intrinsic antiferroelectricity will become accessible in \ce{NbOCl2} monolayer. However, simulation of antiferroelectricity within AFE-\ce{NbOCl2} calls for the accurately predicting kinetic and thermodynamic properties associated with structural transitions between AFE and FE phases \cite{PRB_110_054109_2024,MGE_Advances_3_e70012_2025}, which is beyond the capacity of zero-temperature DFT calculations. The dilemma between computational cost and accuracy can be well addressed by the DP model, where the interatomic forces are accurately trained based on DFT datasets using deep neural network algorithms \cite{DPmodel}. In fact, DP model and associated DPMD simulations can decipher the domain-wall motion and polarization switching kinetics covering the large space and time scale for sliding ferroelectrics \cite{PRL_135_046201_2025,PRB_112_035421_2025}. Therefore, reproducing the complex configuration space and accurately predicting the intrinsic 2D ferroelectricity as well as antiferroelectricity of \ce{NbOCl2} is also feasible by performing large-scale MD simulations based on DP model with quantum accuracy.

\begin{figure}[!htbp]
	\centering
	\includegraphics[width=\linewidth]{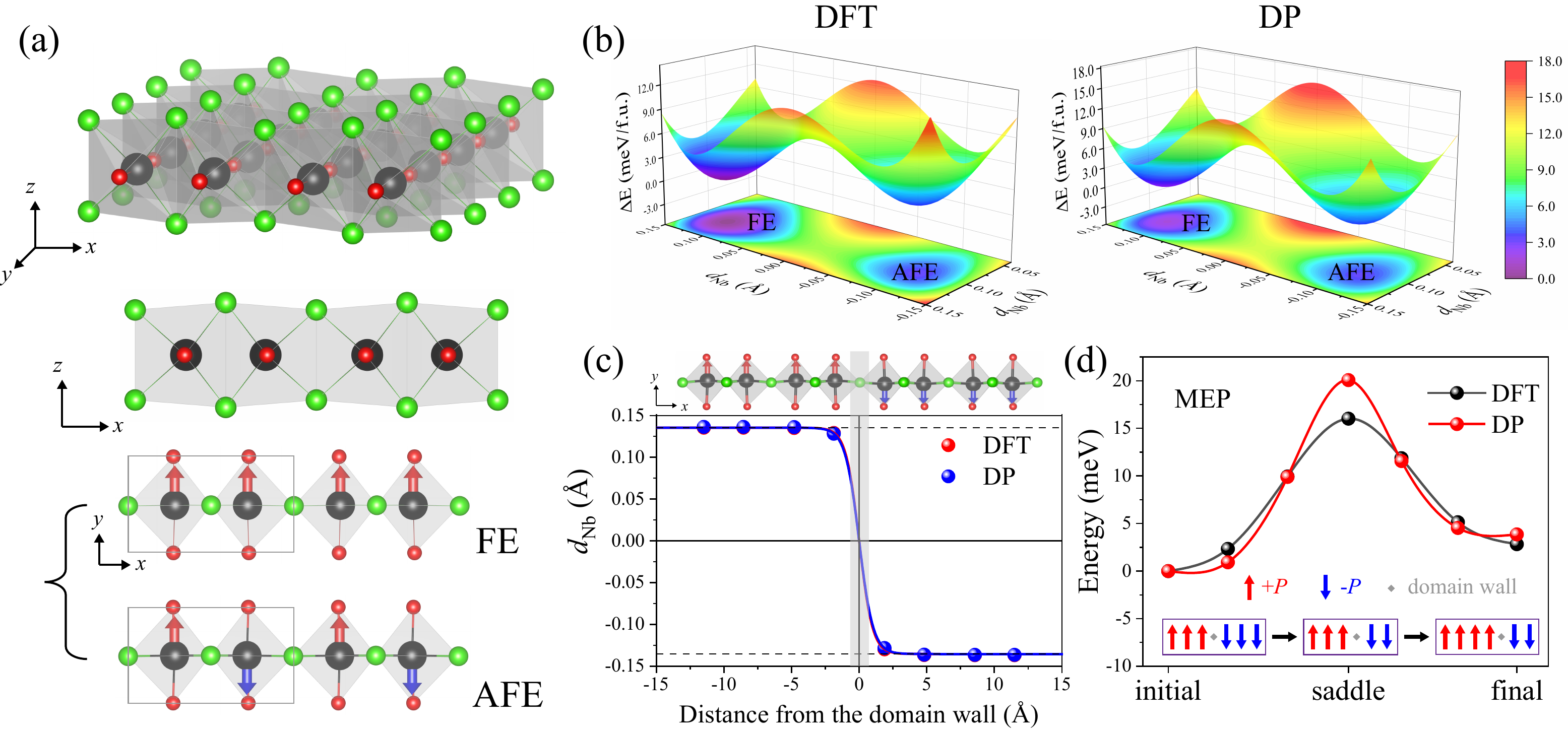}
	\caption{
		(a) Crystal structure for 2D \ce{NbOCl2} monolayer. Nb, O and Cl atoms are in black, red and green colors, respectively. Nb polar displacement in the ground-state stable FE and metastable AFE phases along $\pm y$ direction are indicated by red and blue arrows, respectively. Unit cells of FE- or AFE-\ce{NbOCl2} monolayer are marked as black squareness.
		(b) The potential energy surfaces (PES) (in meV f.u.\textsuperscript{-1}) for relative displacement of two neighboring Nb cations within the free-standing \ce{NbOCl2} monolayer, calculated by DFT and DP model. Both FE and AFE states from PES are identified.
		(c) In the optimized $180^\circ$ FE domain-wall configuration of \ce{NbOCl2} monolayer, variation of $d_\text{Nb}$ as a function of distance ($r$) away from the domain-wall, simulated by DFT (marked in red color) and DP model (in blue) respectively. $180^\circ$ domain-wall with sharp width are in gray area.
		(d) The minimum energy pathway (MEP) and energy barriers for polarization switching within FE-\ce{NbOCl2} monolayer predicted by DFT and DP model. The schematic diagram for domain-wall motion induced polarization switching scenario is also provided.}
		\label{fig1}
\end{figure}

From the crystal structure point of view, 2D vdW layered \ce{NbOCl2} manifests as the structural derivative of \ce{\textit{AB}O3} perovskite without \ce{\textit{A}}-site cation and out-of-plane connectivity, where 2D structural framework of \ce{NbOCl2} monolayer is composed of \ce{NbO2Cl4} octahedrons with mixed corner- and edge-sharing connectivity along two planar directions. As displayed in Fig.~\ref{fig1}(a), Nb-Nb Peierls distortion occurs along Nb-Cl-Nb bonding direction, while polar displacement of Nb cations relative to octahedral centers along Nb-O-Nb direction can lead to the in-plane polarization. Besides, the low-lying AFE phase also presents in \ce{NbOCl2} monolayer via opposite displacement of Nb cations from two neighboring octahedrons (nonequivalent Nb sublattices). FE and AFE phases are determined to be the ground-state and metastable structural phases for free-standing \ce{NbOCl2} monolayer, respectively \cite{Jia2019}, both of which are associated with the zone-center structural instabilities out of the high-symmetry paraelectric (PE) reference.

Owing to the negligible interlayer electronic coupling \cite{Nature_613_53_2023}, the energetic and structural properties of 2D \ce{NbOCl2} are nearly independent on the number of monolayers included, we therefore develop the first-principles-derived DP model \cite{DPmodel,NEURIPS2018,DPGEN} for \ce{NbOCl2} monolayer (see Section~I and Fig.~S1 in Supplementary), to simulate the intrinsic ferroelectricity and potential antiferroelectricity of 2D \ce{NbOCl2}. Our well trained DP model can predict the static properties, such as energies, atomic forces and phonon dispersion in a satisfactory accuracy comparable to DFT results (Section~II and Fig.~S2-S4 in Supplementary). The accuracy of DP model is also validated by reproducing DFT potential energy surface (PES) of free-standing \ce{NbOCl2} monolayer, which refers to the variation of system energy as a function of Nb polar displacement ($d_\text{Nb}$) between two nonequivalent Nb sublattices. Both $d_\text{Nb}$ magnitudes and energetic stability for the stable FE and metastable AFE states within PES are well captured by DP model (Fig.~\ref{fig1}(b)).

Beside the static properties, accuracy in predicting the kinetic process of \ce{NbOCl2} monolayer, such as energy pathways and barriers associated with domain wall motion and polarization switching by DP model have also been evaluated. As $d_\text{Nb}$ is restricted along the one planar axis of but completely forbidden along the other, formation of $180^\circ$ domain-wall between oppositely orientated FE single domains has been predicted \cite{Jia2019} and experimentally validated \cite{ACSNano2023} in 2D \ce{NbOCl2}. Shown in Fig.~\ref{fig1}(c) is variation of $d_\text{Nb}$ as a function of position ($r$) away from the domain wall, where $d_\text{Nb}$-$r$ profiles predicted by DFT and DP model are nearly identical, yielding a sharp domain wall width as small as 2~\AA\ (section~III of Supplementary). Atomically thin domain-wall along Nb-Cl-Nb direction of \ce{NbOCl2} can be rationalized by the flat soft phonon that is nearly dispersionless along the corresponding reciprocal path(Fig.~S4(b) in Supplementary). Similar relation between flat soft phonon and atomically thin boundary width has also been established in hafnia ferroelectrics \cite{HfO2Science2020,HfO2PRL2025}. Specifically, sharp domain-wall width can facilitate the polarization switching kinetics of 2D \ce{NbOCl2} through the motion of domain-wall. As shown in Fig.~\ref{fig1}(d), transition from the initial to final state across saddle point along the simulated minimum energy pathway (MEP) resembles the domain-wall motion induced polarization switching. The MEP associated with polarization switching predicted by DP model based on NEB method agrees well with DFT results, except DP model slightly overestimates the energy barrier by 4~meV. Overall, the DP model can well reproduce the static and kinetic properties of the free-standing \ce{NbOCl2} monolayer obtained from DFT calculations, establishing a solid foundation for accurately predicting its thermodynamic properties at finite temperature.
\begin{figure}[!htbp]
	\centering
	\includegraphics[width=\linewidth]{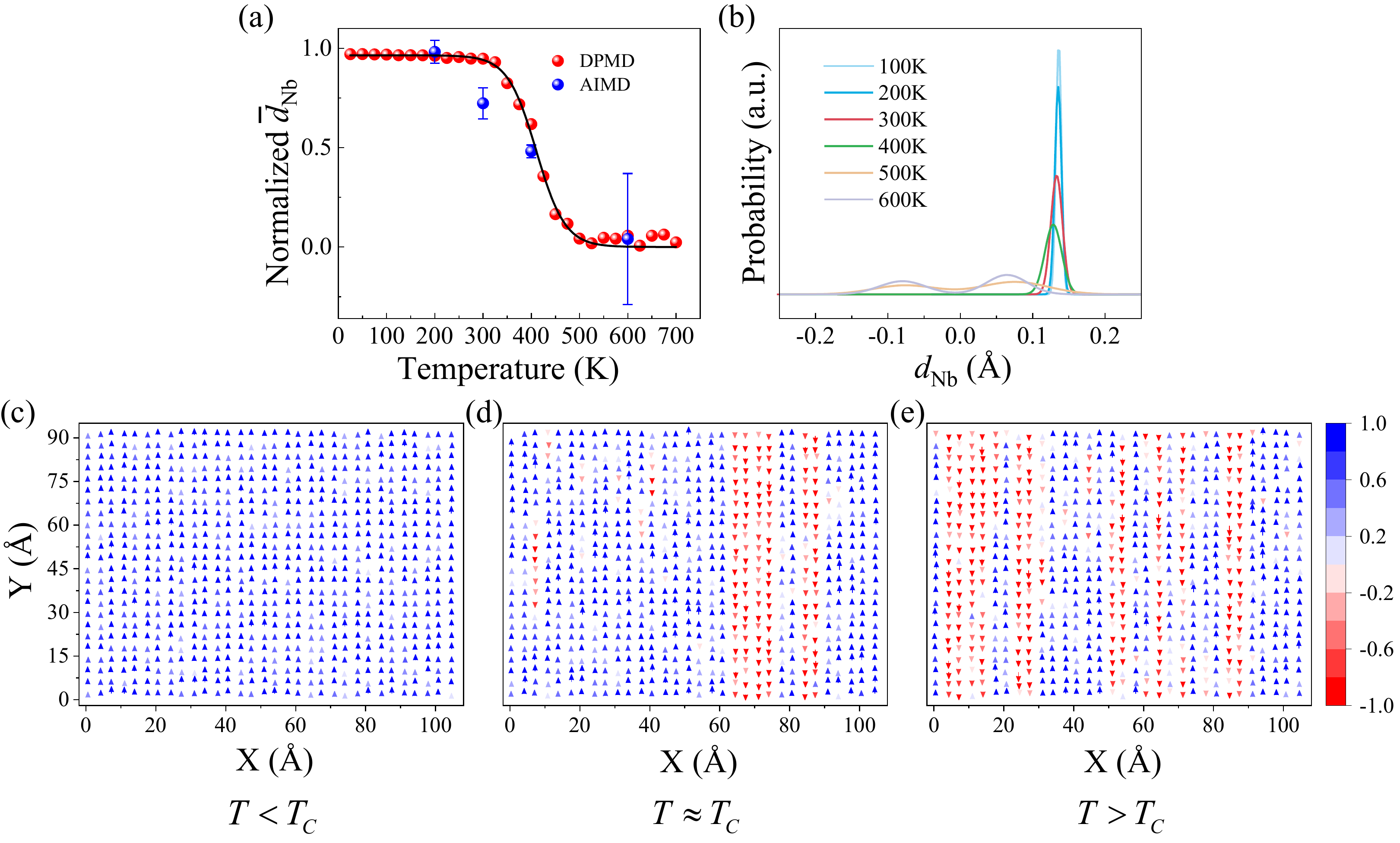}
	\caption{Temperature induced FE-PE phase transition in the free-standing \ce{NbOCl2} monolayer.
	(a) Temperature dependent statistically averaged $\bar{d}_{\mathrm{Nb}}$ normalized to its zero-temperature value, predicted by \textit{ab initio} MD (AIMD) and DPMD simulations respectively. The large error bars reflect the thermal fluctuations during AIMD simulations performed on \ce{NbOCl2} monolayer with small supercell.
	(b) The probability distributions of $d_{\mathrm{Nb}}$ at different temperature.
	(c)-(e) Snapshots of FE-\ce{NbOCl2} monolayer obtained from DPMD simulations performed at temperature below, around and above $T_c$. Each \ce{NbOCl2} monolayer configuration contains $16 \times 24$ Nb octahedrons, where $d_\text{Nb}$ are indicated by colored arrows with different magnitudes. Reversal of polarization orientations are indicated by changing blue arrows into red along the entire Nb lattice columns.
	}
	\label{fig2}
\end{figure}

In experiment, the stable ferroelectricity and strong optical nonlinearly arised from structural noncentrosymmetry have been detected in 2D \ce{NbOCl2} at room temperature \cite{Nature_613_53_2023,AdvOptMat2023,NanoLett.2024,NatComm2023}. However, in-situ experimental characterizations of local polarization with atomic resolution remain challenging, the microscopic mechanism regarding temperature induced phase transition in 2D \ce{NbOCl2} is still unknown. Based on the DP model, we can perform finite-temperature MD simulations of \ce{NbOCl2} at the mesoscopic scale, establishing the relation between macroscopic ferroelectricity and microscopic information (Fig.~\ref{fig2}(a)). The statistically averaged $\bar{d}_{\mathrm{Nb}}$ serves as a order parameter to quantitate the polarization magnitude, while the probability distribution of $d_\text{Nb}$ (Fig.~\ref{fig2}(b)) and real space tracing of $d_\text{Nb}$ on each lattice grid within \ce{NbOCl2} monolayer (Fig.~\ref{fig2}(c)-(e)) as a function of temperature can provide more atomic information regarding the phase transition mechanism. DPMD simulations are performed on FE-\ce{NbOCl2} monolayer with exactly same $d_\text{Nb}$ on each lattice grid (Fig.~\ref{fig2}(c)). FE-to-PE phase transition occurs when temperature approaches to Curie temperature ($T_c$ $\approx$ 396 K), accompanied by sudden drop of $\bar{d}_{\mathrm{Nb}}$ magnitude (Fig.~\ref{fig2}(a)) and shift positions of $d_\text{Nb}$ probability distribution peaks towards the lower values (Fig.~\ref{fig2}(b)), as well as reversing the orientations of $d_\text{Nb}$ for entire column along the polar axis (Fig.~\ref{fig2}(d)). Above $T_c$, \ce{NbOCl2} monolayer enters PE state: each grid has none-zero $d_\text{Nb}$, the overall random orientation of $d_\text{Nb}$ for individual column leads to $\bar{d}_{\mathrm{Nb}}$ = 0 (Fig.~\ref{fig2}(e)). Within PE state, distributions of $d_\text{Nb}$ correspond to two broad probability peaks with opposite peak positions. The magnitudes of $\bar{d}_{\mathrm{Nb}}$ predicted by DPMD simulations are in good agreement with ensemble averaged $\bar{d}_{\mathrm{Nb}}$ from AIMD simulations performed at the selected temperature \cite{Jia2019}. Moreover, compared to the large thermal fluctuations during AIMD simulations, the more statistically reliable results can be produced by DPMD simulations performed on ultra-large supercell to eliminate the supercell size effect \cite{SCsize,prbIn2Se3/size}.

\begin{figure}[!htbp]
	\centering
	\includegraphics[width=0.8\linewidth]{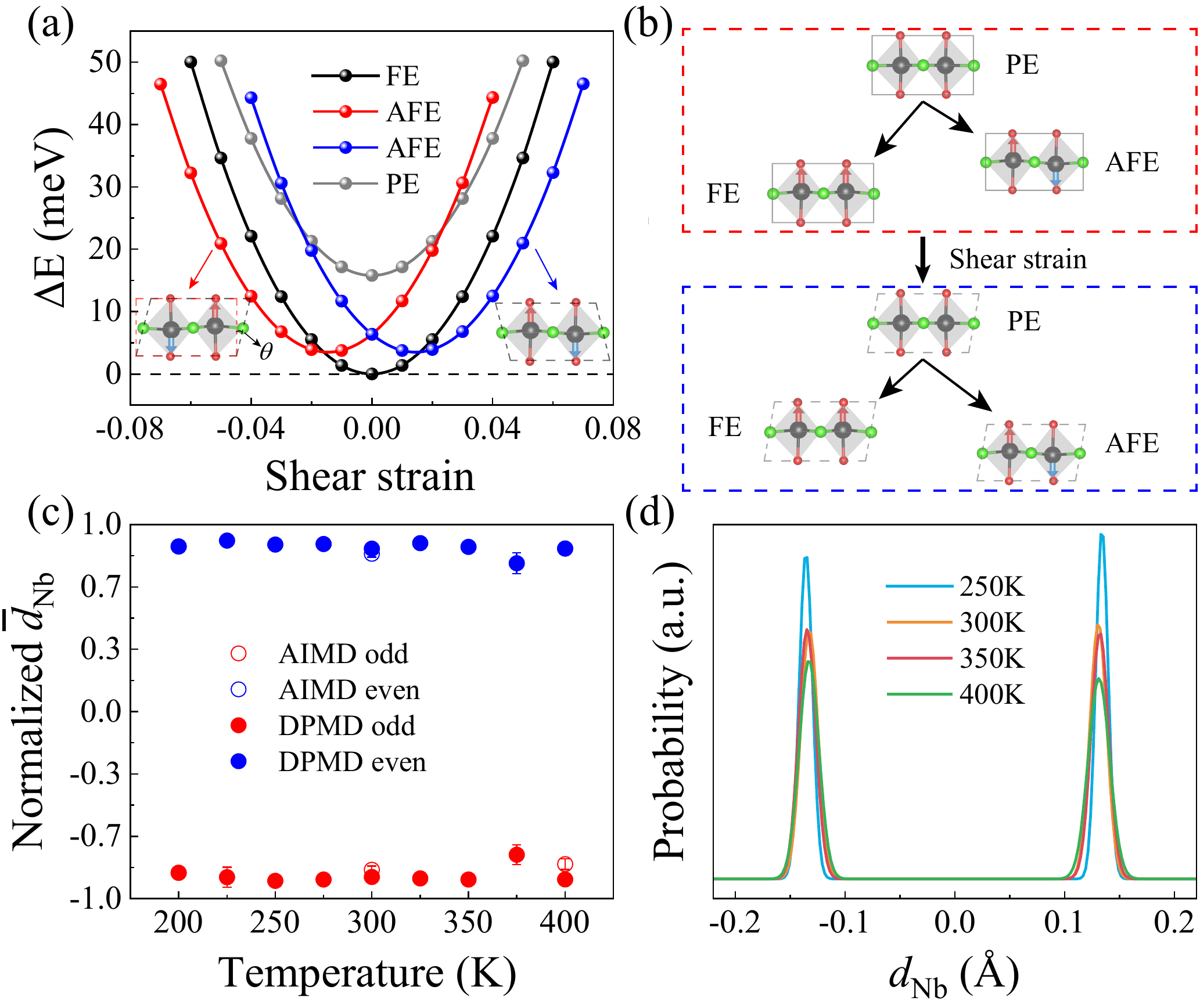}
	\caption{
	(a) The variation of system energies for \ce{NbOCl2} monolayer in FE, AFE and PE phases as a function of the applied shear strain, predicted by DFT calculations. The energy of free-standing (strain free) FE-\ce{NbOCl2} monolayer is set as energy zero. Inset illustrates how 2D lattices of \ce{NbOCl2} monolayer are deformed under shear strain.
	(b) Schematic diagram for stabilization of antiferroelectricity within \ce{NbOCl2} monolayer by the shear strain, where the previous metastable AFE phase becomes the ground-state stable phase when \ce{NbOCl2} monolayer is under the certain shear strain.
	(c) Variantion of the normalized $\bar{d}_{\mathrm{Nb}}$ from from two inequivalent Nb sublattices (odd and even Nb lattice columns) within AFE-\ce{NbOCl2} monolayer as a function of temperature, predicted by AIMD and DPMD simulations.
	(d) Probability distributions of $d_\text{Nb}$ within AFE-\ce{NbOCl2} monolayer at different temperature. In AFE-\ce{NbOCl2} monolayer, $\bar{d}_{\mathrm{Nb}}$ from two nonequivalent Nb sublattices are in nearly same magnitudes, but with opposite signs.}
	\label{fig3}
\end{figure}
Based on zero temperature DFT calculations for free-standing \ce{NbOCl2} monolayer, the metastable AFE phase is very close to FE phase in energy ($\Delta \text{E}$ $\sim$ 2 meV f.u.\textsuperscript{-1}), but antiferroelectricity has never been detected in 2D \ce{NbOCl2} by experiment. In fact, even we perform finite-temperature DPMD simulations on AFE-\ce{NbOCl2} monolayer, AFE phase still cannot be stabilized at room temperature. Instead, \ce{NbOCl2} monolayer adopts FE state with the oppositely orientated FE single domains and nonzero $\bar{d}_{\mathrm{Nb}}$ magnitude (Fig.~S5 in Supplementary). Therefore, the alternative control parameter rather than temperature should be employed to stabilize AFE phase of \ce{NbOCl2}.

The proper control parameter (e.g. compositional substitution or elastic strain) that can alter the energy sequence of FE and AFE phases, may also trigger the structural transition between the two phases \cite{KarinRabe}. As for 2D \ce{NbOCl2}, the intrinsic antiferroelectricity can be stabilized by experimentally accessible shear strain. Fig.~\ref{fig3}(a) displays the variation of system energies for FE, AFE and PE phases of \ce{NbOCl2} monolayer as a function of the applied shear strain, whose magnitude refers to the radian value of the shear strain induced angle change between two planar axes of \ce{NbOCl2} monolayer. Without shear strain, free-standing \ce{NbOCl2} monolayer has strong preference towards the ground-state FE phase. The alternation of energy sequence between FE and AFE phases occurs when the magnitude of shear strain applied \ce{NbOCl2} monolayer is above 2\%, which is readily achievable by experiment \cite{MoS22021}. Strain induced FE-AFE phase transition becomes more apparent in schematic diagram displayed in Fig.~\ref{fig3}(b). Owing to the small energy difference and delicate competition between FE and AFE phases \cite{KarinRabe}, shear strain in moderate magnitude can alter the energy preference of 2D \ce{NbOCl2}, making the previous metastable AFE structure as the ground-state stable phase.

Above prediction regarding AFE-\ce{NbOCl2} monolayer under shear strain are made by zero temperature DFT calculations. In order to access intrinsic antiferroelectricity from 2D \ce{NbOCl2} by experiment, the thermal stability of AFE-\ce{NbOCl2} should be evaluated by finite temperature MD simulations. To this end, following the similar procedure for free-standing \ce{NbOCl2} monolayer, we also develop the DP model for \ce{NbOCl2} monolayer under 5\% shear strain (See Fig.~S6 in Supplementary). Finite temperature properties of AFE-\ce{NbOCl2} are then examined by performing DPMD simulations on strained \ce{NbOCl2} monolayer. Different from FE phase, $\bar{d}_{\mathrm{Nb}}$ from two nonequivalent Nb sublattices (odd and even Nb octahedral columns) are employed as the structural order to quantitatively describe AFE phase. AFE-\ce{NbOCl2} monolayer characterizes the collinear and antiparallel arrangement of $d_\text{Nb}$ between two nonequivalent Nb sublattices ($\downarrow \uparrow \downarrow \uparrow$, see Fig.~S7 in Supplementary), so that statistically averaged $\bar{d}_{\mathrm{Nb}}$ from odd and even columns are in nearly same magnitudes but with opposite signs (Fig.~\ref{fig3}(c)), leading to two symmetric $d_\text{Nb}$ probability distribution peaks (Fig.~\ref{fig3}(d)). Moreover, antiferroelectricity in strained \ce{NbOCl2} monolayer can persist above room temperature, where $\left| \bar{d}_{\mathrm{Nb}} \right|$ at 300~K decreases slightly from the low temperature values and shift positions of $d_\text{Nb}$ probability peaks towards the lower amplitudes. Once again, $\bar{d}_{\mathrm{Nb}}$ predicted by DPMD simulations are nearly coincident with ensemble averaged $\bar{d}_{\mathrm{Nb}}$ obtained from AIMD simulations, which convincingly demonstrates the room temperature stable antiferroelectricity in strained \ce{NbOCl2} monolayer.
\begin{figure}[!htbp]
	\centering
	\includegraphics[width=0.9\linewidth]{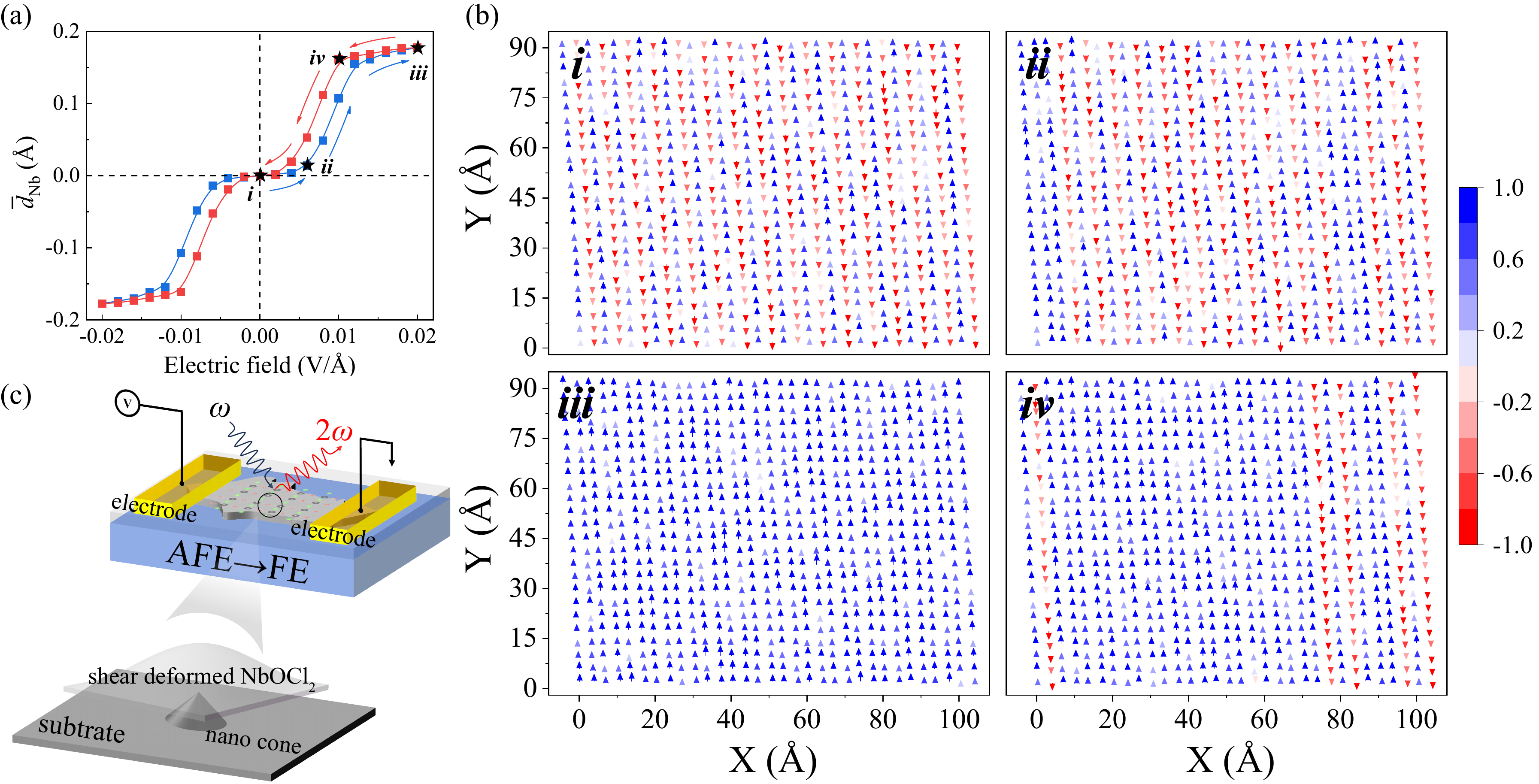}
	\caption{Electric field induced AFE-to-FE phase transition in \ce{NbOCl2} monolayer under 5\% shear strain.
	(a) Double hysteresis loop by recording the variation of statistically averaged $\bar{d}_{\mathrm{Nb}}$ as a function electric field $\mathcal{E}$ applied along polar $y$ axis, obtained from DPMD simulations performed around room temperature.
	(b) Tracing of $d_\text{Nb}$ within \ce{NbOCl2} monolayer for four prototypical snapshot configurations along the hysteresis loop. At stage $i$ and $iii$, strained-\ce{NbOCl2} monolayer is in AFE and FE phases, respectively. Reversible transition between FE and AFE phases occur at state $ii$ and $iv$, when \ce{NbOCl2} is under the low critical field ($\mathcal{E}_c$ $<$ 0.01 V/\AA).
	(c) Schematic diagram for electric control of optical nonlinearity in 2D \ce{NbOCl2}. Tuning the centrosymmetric AFE phase free of SHG signal into FE phase with giant SHG response can be realized in shear deformed \ce{NbOCl2} nanoflake by electric poling.}
	\label{fig4}
\end{figure}

Beside the structural characters, simulating the electric field induced transition from AFE to FE phase can provide the most definitive indication of antiferroelectricity \cite{KarinRabe}. Thanks to the predictive capacity of DP model, we can simulate hysteresis loop of AFE-\ce{NbOCl2} monolayer under the applied electric filed that can directly resemble the experimental measurement. Shown in Fig.~\ref{fig4}(a) is our simulated hysteresis loop by recording change of $\bar{d}_{\mathrm{Nb}}$ with respect to electric field $\mathcal{E}$ applied along polar $y$ axis of strained \ce{NbOCl2} monolayer, obtained from DPMD simulations performed around room temperature. The simulated double hysteresis loop of strained \ce{NbOCl2} monolayer is the characteristic feature of intrinsic antiferroelectricity. Without external electric field, strained \ce{NbOCl2} monolayer crystallizes in AFE phase (stage $i$ of Fig.~\ref{fig4}(b)). $\bar{d}_{\mathrm{Nb}}$ magnitude increases gradually when \ce{NbOCl2} is under positive electric poling. AFE-to-FE transition occurs at stage $ii$, where $d_\text{Nb}$ along the entire lattice column reverse their directions. When electric field is above 0.016 V/\AA, \ce{NbOCl2} enters FE state with saturated polarization (stage $iii$). The reversible FE-to-AFE phase transition occurs at stage $iv$. Two critical electric fields responsible for AFE-to-FE and FE-to-AFE transitions are close in magnitude ($\mathcal{E}_\text{AFE-FE}$ = 0.006 $vs$ $\mathcal{E}_\text{FE-AFE}$ = 0.01 V/\AA), leading to $P$-$E$ loop of small hysteresis. It is noted that $P$-$E$ loops with more obvious hysteresis are predicted by DPMD simulations performed at lower temperature (Fig.~S8 in Supplementary). Therefore, the overall small hysteresis and low critical electric fields in AFE-\ce{NbOCl2} around room temperature are associated with small polarization switching barrier to be overcomed by the external electric field \cite{Xu2017}.

In experiment, rich research efforts have been dedicated to engineer SHG responses of \ce{NbOCl2} \cite{NanoLett.2024,NatComm2023,APL2025}. However, electric control of the optical nonlinearity in 2D \ce{NbOCl2} has never been realized. After switching for polarization of FE-\ce{NbOCl2} by external electric field, $\pm P$ states with opposite polarization orientations characterize the same SHG responses \cite{Nature_613_53_2023}. With emerging of intrinsic antiferroelectricity, new scenario for engineering the optical nonlinearity of \ce{NbOCl2} can be realized (Fig.~\ref{fig4}(c)). As-prepared \ce{NbOCl2} nanoflakes can be deposited on a hard nanocone by experiment. The strain can be transferred from hard nanocone to \ce{NbOCl2} by interfacial shear deformation \cite{AMnanocone2019}. The amplitude of effective shear strain on \ce{NbOCl2} can be determined by in-situ Raman characterization \cite{MoS22021}. Antiferroelectricity appears when \ce{NbOCl2} nanoflakes are under certain deformation. Transition from the centrosymmetric AFE phase free of SHG signal into FE-\ce{NbOCl2} with giant SHG response by electric poling, enables the electric control of second-order optical nonlinearity in 2D \ce{NbOCl2}. Moreover, compared to the experimentally synthesized 2D AFE materials (e.g. $\alpha$-GeSe \cite{ACS_Nano_16_1308_2022} and \ce{CuBiP2Se6} \cite{Adv_Mater_37_2419204_2025}), antiferroelectricity of strained \ce{NbOCl2} is able to persist up to monolayer limit. Therefore, combing the small hysteresis and low critical field during AFE-to-FE phase transition, electric "writing" and nonlinear optical "reading" in optoelectronic device composed of AFE-\ce{NbOCl2} with low power consumption and fast operation speed can be expected.

In summary, first-principles-derived interatomic potential were developed for 2D \ce{NbOCl2} with intrinsic ferroelectricity and antiferroelectricity. Large-scale DPMD simulations were performed to reveal the macroscopic properties and microscopic mechanisms regarding the temperature and electric field driven phase transitions of \ce{NbOCl2}. For free-standing \ce{NbOCl2} monolayer with in-plane ferroelectricity, DP model can well predict its energy landscape, atomically sharp domain wall and Curie temperature with sufficient accuracy. Beside the well studied 2D ferroelectricity, accessing the previously overlooked antiferroelectricity in \ce{NbOCl2} via strain engineering approach has also been demonstrated. The room temperature stable antiferroelectricity was predicted in 2D \ce{NbOCl2} monolayer under the experimentally accessible shear strain, according to MD simulations based on the iteratively refined DP model. Especially, owing to the unique collinear arrangement of local polarizations restricted along one planar axis and low polarization switching barrier, transition from AFE phase free of SHG response to FE phase with giant SHG intensity can be triggered by low critical electric field, rendering the double $P$-$E$ loop with small hysteresis. Based on the multiscale modeling performed on \ce{NbOCl2} with competing FE and AFE ordering, our work not only updates the current understanding regarding 2D antiferroelectricity, but also provides a new strategy for electric control of optical nonlinearity in \ce{NbOCl2} via electric field induced AFE-to-FE phase transition.

\begin{acknowledgments}
	G.Y.G. acknowledges the funding support from the National Natural Science Foundation of China (Grant No. 11574244). X.C.Z. acknowledges the support from the Hong Kong Global STEM Professorship Scheme. Prof. Bin Xu from Soochow University is acknowledged for helpful discussion. Hefei Advanced Computing Center is acknowledged for computational support.
\end{acknowledgments}

\bibliography{refs}
\end{document}